\documentclass[fp]{jpsj3}
\usepackage{txfonts,tabularx}
\usepackage[usenames]{color}

\begin{document}

\title{Doping Dependencies of Onset Temperatures for the Pseudogap and Superconductive Fluctuation in Bi$_{2}$Sr$_{2}$CaCu$_{2}$O$_{8+\delta}$, Studied from both In-Plane and Out-of-Plane Magnetoresistance Measurements}

\author{\name{Tomohiro \surname{Usui}}$^1$\thanks{E-mail: h12gs702@stu.hirosaki-u.ac.jp}, \name{Daiki \surname{Fujiwara}}$^1$, \name{Shintaro \surname{Adachi}}$^1$, \name{Hironobu \surname{Kudo}}$^1$, \name{Kosuke \surname{Murata}}$^1$, \name{Haruki \surname{Kushibiki}}$^1$, \name{Takao \surname{Watanabe}}$^1$, \name{Kazutaka \surname{Kudo}}$^2$\thanks{Present address: Department of Physics, Okayama University, 3-1-1 Tsushima-naka, Okayama 700-8530, Japan}, \name{Terukazu \surname{Nishizaki}}$^2$\thanks{Present address: Department of Electrical Engineering and Information Technology, Kyushu Sangyo University, Fukuoka 813-8503, Japan}, \name{Norio \surname{Kobayashi}}$^2$, \name{Shojiro \surname{Kimura}}$^2$, \name{Kazuyoshi \surname{Yamada}}$^2$\thanks{Present address: Institute of Materials Structure Science, KEK, Tsukuba, Ibaraki, 305-0801, Japan}, \name{Tomoyuki \surname{Naito}}$^3$, \name{Takashi \surname{Noji}}$^4$, and \name{Yoji \surname{Koike}}$^4$}
\inst{$^1$Graduate School of Science and Technology, Hirosaki University, 3 Bunkyo, Hirosaki, 036-8561 Japan \\
$^2$Institute for Materials Research, Tohoku University, 2-1-1 Katahira, Aoba-ku, Sendai, 980-8577 Japan \\
$^3$Faculty of Engineering, Iwate University, 4-3-5 Ueda, Morioka, 020-8551 Japan \\
$^4$Graduate School of Engineering, Tohoku University, 6-6-05 Aoba, Aramaki, Aoba-ku, Sendai, 980-8579 Japan} 

\abst{To investigate the relationship between the pseudogap and superconductivity, we measured both the in-plane ($\rho_{ab}$) and out-of-plane ($\rho_c$) resistivity for oxygen-controlled Bi$_{2}$Sr$_{2}$CaCu$_{2}$O$_{8+\delta}$ single crystals subject to magnetic fields (parallel to the $c$ axis) of up to 17.5 T. The onset temperature for the superconductive fluctuation, $T_{scf}$, is determined by the large positive in-plane magnetoresistance (MR) and negative out-of-plane MR observed near $T_c$, whereas the pseudogap opening temperature $T^*$ is determined by the semiconductive upturn of the zero-field $\rho_c$. $T_{scf}$ was found to scale roughly as $T_c$, with a decreasing temperature interval between them upon doping. On the other hand, $T^*$ starts out much higher than $T_{scf}$ but decreases monotonically upon doping; finally, at a heavily overdoped state, it is not observed above $T_{scf}$. These results imply that the pseudogap is not a simple precursor of superconductivity, but that further study is needed to determine whether or not $T^*$ exists below $T_{scf}$ in the heavily overdoped state.}

\kword{superconductivity, pseudogap, superconductive fluctuation, doping, magnetoresistance, Bi-2212}

\maketitle

\section{INTRODUCTION}

It is well recognized that knowledge of the pseudogap phenomena in high-$T_c$ cuprates is crucial to understand the mechanism of high-$T_c$ superconductivity. It is not yet conclusive whether it is a necessary ingredient (precursor) for high-$T_c$ superconductivity~\cite{emery,suz,lee,huf,chat} or if it is a different ordered state~\cite{var,tanak,lawl}. In the former case, the pseudogap opening temperature $T^*$ coincides with the onset temperature of superconductive fluctuation, $T_{scf}$, and may merge with $T_c$ in the overdoped state. In the latter case, the $T^*$ behavior is independent from $T_{scf}$, and may cross it on its way to zero temperature at the quantum critical point near the optimal doping of $p$ = 0.19~\cite{tal}. Thus, in order to elucidate the role of the pseudogap in high-$T_c$ superconductivity, it is important to investigate the relationship between $T^*$ and $T_{scf}$ as a function of doping. 

Among the various techniques to address this issue, the out-of-plane resistivity $\rho_{c}$ is one of the most powerful methods, since it directly probes the electronic density-of-states (DOS) around the Fermi level, reflecting the tunneling nature between CuO$_{2}$ planes in high-$T_c$ cuprates. Indeed, $\rho_{c}$ shows a typical upturn below a characteristic temperature $T^*_{\rho_c}$~\cite{wata2,t.shiba,lav,take}. This means that the DOS around the Fermi level is decreased by the opening of some kind of a gap below $T^*_{\rho_c}$. The $T^*_{\rho_c}$ has been found to coincide well with $T^*_{\chi}$~\cite{wata2}, below which magnetic susceptibility gradually decreases. Since magnetic susceptibility also reflects the DOS (Pauli paramagnetism), the coincidence of $T^*_{\rho_{c}}$ and $T^*_{\chi}$ simply indicates that these temperatures are the so called pseudogap opening temperature $T^*$. Tunneling measurements also support this view~\cite{matsuda,suzuki,kra1}. Furthermore, the $\rho_{c}$ of Bi$_{2}$Sr$_{2}$CaCu$_{2}$O$_{8+\delta}$ (Bi-2212) has been measured in high magnetic fields up to 60 T, and a negative magnetoresistance (MR) was observed below $T^{*}$~\cite{t.shiba}. This result has been attributed to the recovery of the DOS that was suppressed in the pseudogap state. Accordingly, we can safely estimate $T^*$ by the onset of the upturn in $\rho_{c}$ measurements.

On the other hand, in addition to a small out-of-plane negative MR below $T^{*}$, rapid growth near $T_c$ has been observed in Bi$_{2+z}$Sr$_{2-x-z}$La$_{x}$CuO$_{y}$ (Bi-2201)~\cite{lav}, Pb-substituted Bi-2201~\cite{kudo}, and in slightly overdoped Bi-2212~\cite{heine}. In close relation to these observations, an out-of-plane negative MR in the superconducting state has been observed for Bi-2212~\cite{moro} by suppressing the Josephson current. Intrinsic tunneling spectroscopy has also revealed that the pseudogap is insensitive to magnetic fields while the superconducting gap is sensitive to it~\cite{kra}. All these results suggest that the large out-of-plane negative MR is caused by the opening of the superconducting gap. Indeed, a theoretical paper~\cite{varla} proposes that in highly anisotropic materials such as high-$T_c$ cuprates, the superconductive DOS fluctuation effect dominates at a wide temperature region above $T_c$, causing the upturn and thus the negative MR in $\rho_{c}$. In extremely two-dimensional (2D) systems, the Aslamasov-Larkin (AL) contribution, which describes the paraconductive contribution (positive MR) by a short-lived superconductive droplet in thermal nonequilibrium, is considerably suppressed for $\rho_{c}$, since the AL contribution is proportional to the square of the small probability of out-of-plane electron hopping $p_1$ ($\sigma_{c}^{AL}\propto p_1^2/\varepsilon^2$, where $\varepsilon=\ln(T/T_c)$). Thus, the AL contribution is important only in close vicinity to $T_c$. In turn, the DOS contribution, which is the reduction of the quasi-particle DOS by the opening of the superconducting gap causing an increase in $\rho_{c}$ (negative MR), dominates in a wide temperature region above $T_c$. This is due to the fact that the DOS contribution depends only linearly on the out-of-plane electron hopping $p_1$ and DOS recovers from superconducting to normal state very slowly in the 2D case ($\sigma_{c}^{DOS}\propto -p_1\ln(1/\varepsilon)$)~\cite{varla}. The Maki-Thompson (MT) contribution may be very small in high-$T_c$ cuprates due to strong pair-breaking effects~\cite{semba}. Thus, as temperature increases, a crossover from AL to DOS fluctuation takes place~\cite{varla}. This may cause a sharp increase in $\rho_{c}$ and a large negative out-of-plane MR above $T_c$. Therefore, it is expected that we can estimate $T_{scf}$ from a rapid increase in the negative MR of $\rho_{c}$. However, this method for estimating $T_{scf}$ has not been established previously. Moreover, when the doping level is increased to the overdoped level, $T^{*}$ and $T_{scf}$ come close to each other and thus it is difficult to distinguish between
them. Another way to estimate $T_{scf}$ is to observe a positive MR in the in-plane resistivity $\rho_{ab}$. In the in-plane cases as well, there may be two competing factors: positive AL and negative DOS. It is expected, however, that the positive AL contribution is much larger compared to the out-of-plane cases, since, in the in-plane case, there is no pre-factor such as the out-of-plane hopping probability~\cite{varla}. Thus, we assume that in-plane MR is dominated by the contribution from positive AL just as for other, ordinary, superconductors. Indeed, in-plane MR in high-$T_c$ cuprates usually has been observed to be positive~\cite{heine,wataB}. Therefore, the simultaneous measurements of both $\rho_{ab}$ and $\rho_{c}$ in magnetic fields would enable us to determine $T_{scf}$ in a more accurate way.

In this paper, in order to investigate the relation between the pseudogap and superconductivity, we have systematically measured both $\rho_{ab}$ and $\rho_{c}$ of variously doped Bi-2212 with and without magnetic fields. For this purpose, we successfully prepared a heavily overdoped crystal ($p$ = 0.23, $T_{c}$ = 53 K) along with other variously doped samples. The details of the sample preparation are described in Sec. 2, and the results and discussion of the measurements are presented in Sec. 3. In the latter section, we reconsider our earlier assertions~\cite{wataC}. The implications of the obtained results will be discussed. Finally, the work is summarized in Sec. 4. 

\section{EXPERIMENT}

Single crystals of Bi$_{2+x}$Sr$_{2-x}$CaCu$_{2}$O$_{8+\delta}$ (x = 0, 0.1 and 0.2) and Pb-doped Bi$_{1.6}$Pb$_{0.4}$Sr$_{2}$CaCu$_{2}$O$_{8+\delta}$ were grown in air using the traveling solvent floating zone method. Underdoped samples ($p$ = 0.11) were then made by annealing the crystals with x = 0.2 under appropriate oxygen partial pressures~\cite{wata1}. Optimal ($p$ = 0.16) and slightly overdoped ($p$ = 0.19-0.20) samples were similarly prepared using the crystals with x = 0.1. The Pb-doped crystals were annealed in oxygen flow at 400$^\circ$C for 50 h to make overdoped samples ($p$ = 0.22). Heavily overdoped samples were made by annealing the crystal with x = 0 at 400$^\circ$C for 30 h under high O$_{2}$ pressure (400 atm). The doping level ($p$) was obtained using the empirical relation proposed by Tallon~\cite{tallon}, with maximum $T_c=$ 83, 89, 91, and 93 K for x = 0.2, 0.1, 0, and the Pb-doped samples, respectively. The difference in the values of the maximum $T_c$ may originate in differences in the disorder at the Sr-site~\cite{hobou}. The maximum $T_c$ contains an error of several kelvins so that $p$ contains an error less than 0.005. Here, we have assumed that the empirical relation holds for all the samples examined, although the superconducting region of the Pb-substituted Bi-2201 has been reported to be rather narrow compared with other Bi-based high-$T_c$ cuprates~\cite{kudo1}. The superconducting transition temperature $T_c$ was determined by the onset of zero resistivity.
The $\rho_{ab}$ and $\rho_{c}$ measurements were carried out using a DC four-terminal method. For the $\rho_{c}$ measurements, voltage contacts were attached to the center of the {\it ab} plane, and the current contacts covered almost all the remaining space~\cite{motohashi}. 
Magnetic fields $B$ of up to 17.5 T were applied parallel to the $c$ axis with a superconducting magnet. We used a Cernox resistive sensor, whose magnetic-field-induced temperature reading errors $\Delta T/T$ were less than 0.2$\%$ at temperatures above 20 K under magnetic fields of less than 19 T. Therefore, the errors were ignored in this study.   

\section{RESULTS AND DISCUSSION}

\begin{figure*}
\includegraphics[width=180mm]{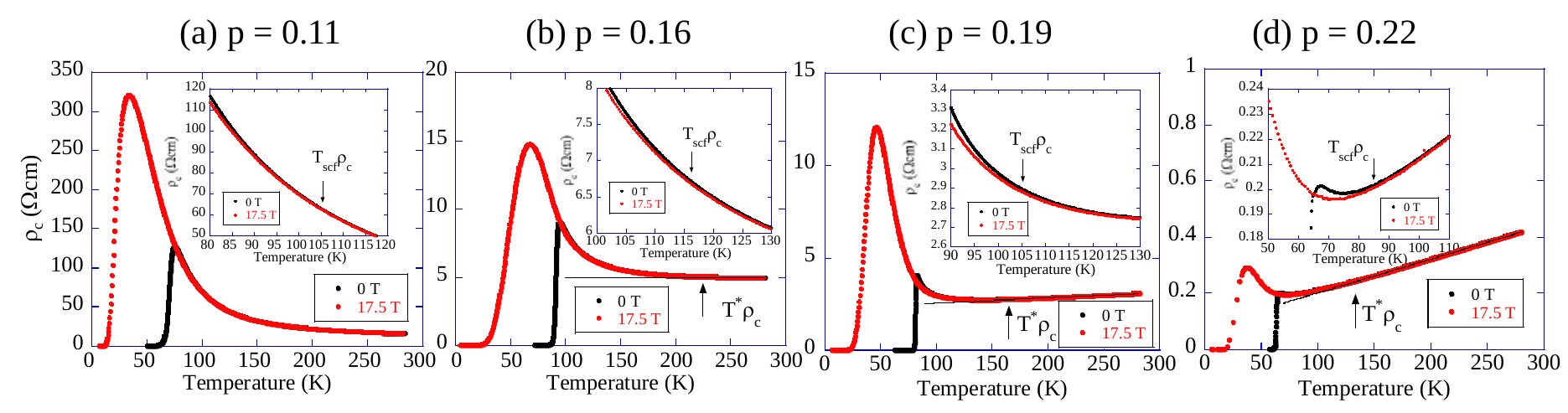}
\caption{\label{f1} (Color online) Out-of-plane resistivity $\rho_c(T)$ for (a) underdoped ($p$ = 0.11), (b) optimally doped ($p$ = 0.16), (c) slightly overdoped ($p$ = 0.19), and (d) overdoped ($p$ = 0.22) Bi$_{2}$Sr$_{2}$CaCu$_{2}$O$_{8+\delta}$ single crystals with and without a magnetic field of 17.5 T. The solid straight lines, which are linear extrapolations of $\rho_c$ at higher temperatures, are included to guide the eye for the (b) optimally, (c) slightly overdoped, and (d) heavily overdoped samples. Arrows indicate the temperatures $T^*_{\rho_c}$ below which $\rho_c$ shows a characteristic upturn. Insets show the expanded view just above $T_c$. Arrows in the insets indicate the temperatures $T_{scf}{\rho_{c}}$, below which $\rho_c$ shows a marked negative magnetoresistance (MR).}
\end{figure*}

Figure \ref{f1} (a)-(d) shows the temperature dependence of the out-of-plane resistivities $\rho_{c}$ for variously doped samples with and without a magnetic field of 17.5 T. At a zero field, the $\rho_{c}$ of an underdoped state ($p$ = 0.11) is semiconductive in all temperature regions measured, implying that the pseudogap opened above room temperature. With increasing doping, the typical upturn behavior below a certain temperature $T^*_{\rho_{c}}$, which is an indication of opening of the pseudogap, is observed~\cite{wata2}. The pseudogap temperature $T^*_{\rho_{c}}$ is estimated as 222, 163 and 136 K for $p$ = 0.16, $p$ = 0.19 and $p$ = 0.22, respectively. Here, $T^*_{\rho_{c}}$  is estimated as the temperature at which $\rho_{c}$ increases 1\% from the high temperature linear behavior. When we apply the magnetic field, we observe a marked negative MR at low temperatures near $T_{c}$. The insets of Fig. \ref{f1} show the enlarged views just above $T_{c}$. The onset temperatures for the rapid growth of negative MR (denoted as $T_{scf}{\rho_{c}}$ in the insets of Fig. \ref{f1}) will be determined below.

\begin{figure*}
\includegraphics[width=180mm]{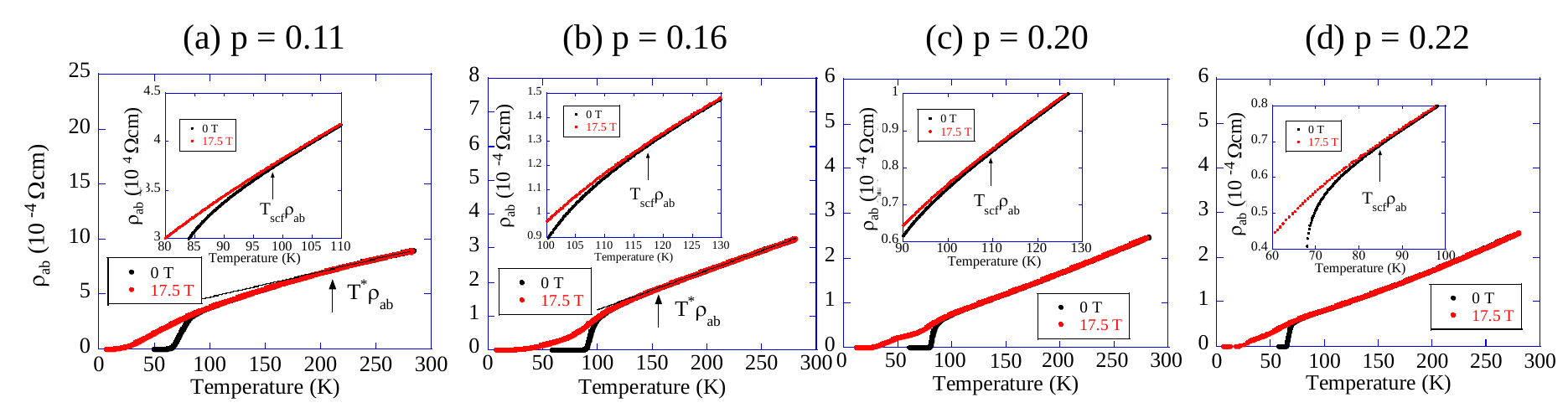}
\caption{\label{f2} (Color online) In-plane resistivity $\rho_{ab}(T)$ for (a) underdoped ($p$ = 0.11), (b) optimally doped ($p$ = 0.16), (c) slightly overdoped ($p$ = 0.20), and (d) overdoped ($p$ = 0.22) Bi$_{2}$Sr$_{2}$CaCu$_{2}$O$_{8+\delta}$ single crystals with and without a magnetic field of 17.5 T. The solid straight lines, which are linear extrapolations of $\rho_{ab}$ at higher temperatures, are drawn to guide the eye for the (a) underdoped and (b) optimally doped samples. The temperatures $T^*_{\rho_{ab}}$ at which $\rho_{ab}$ deviates from linear behavior are shown with arrows. Insets show the expanded view just above $T_c$. Arrows in the insets indicate the temperatures $T_{scf}{\rho_{ab}}$, below which $\rho_{ab}$ shows a marked positive magnetoresistance (MR).}
\end{figure*}

Figure \ref{f2} (a)-(d) shows the temperature dependence of the in-plane resistivities $\rho_{ab}$ for variously doped samples with and without a magnetic field of 17.5 T. At a zero field, the underdoped ($p$ = 0.11) and optimally doped ($p$ = 0.16) samples show a typical downward deviation from high-temperature linear behavior below a certain temperature $T^*_{\rho_{ab}}$, as reported in our previous papers~\cite{wata1,t.fujii}. $T^*_{\rho_{ab}}$ is estimated to be 209 and 153 K for $p$ = 0.11 and $p$ = 0.16, respectively. Here, $T^*_{\rho_{ab}}$  is estimated as the temperature at which $\rho_{ab}$ decreases 1\% from the high temperature linear behavior. It should be noted that the obtained $T^*_{\rho_{ab}}$ agrees very well with the estimated $T^*_{\rho_{ab}}$ when using a different method, {\it i.e.}, resistivity curvature mapping (RCM)~\cite{ando}. By further increased doping, $T^*_{\rho_{ab}}$ becomes difficult to determine and instead an upward curvature appears, indicating that the system enters into the overdoped region~\cite{wata1}. When we apply a magnetic field, we observe a marked positive MR near $T_{c}$. The enlarged views just above $T_{c}$ are shown in the insets of Fig. \ref{f2}. The onset temperatures for the rapid growth of positive MR (denoted as $T_{scf}{\rho_{ab}}$ in the insets of Fig. \ref{f2}) will be determined below.

\begin{figure}
\includegraphics[width=85mm]{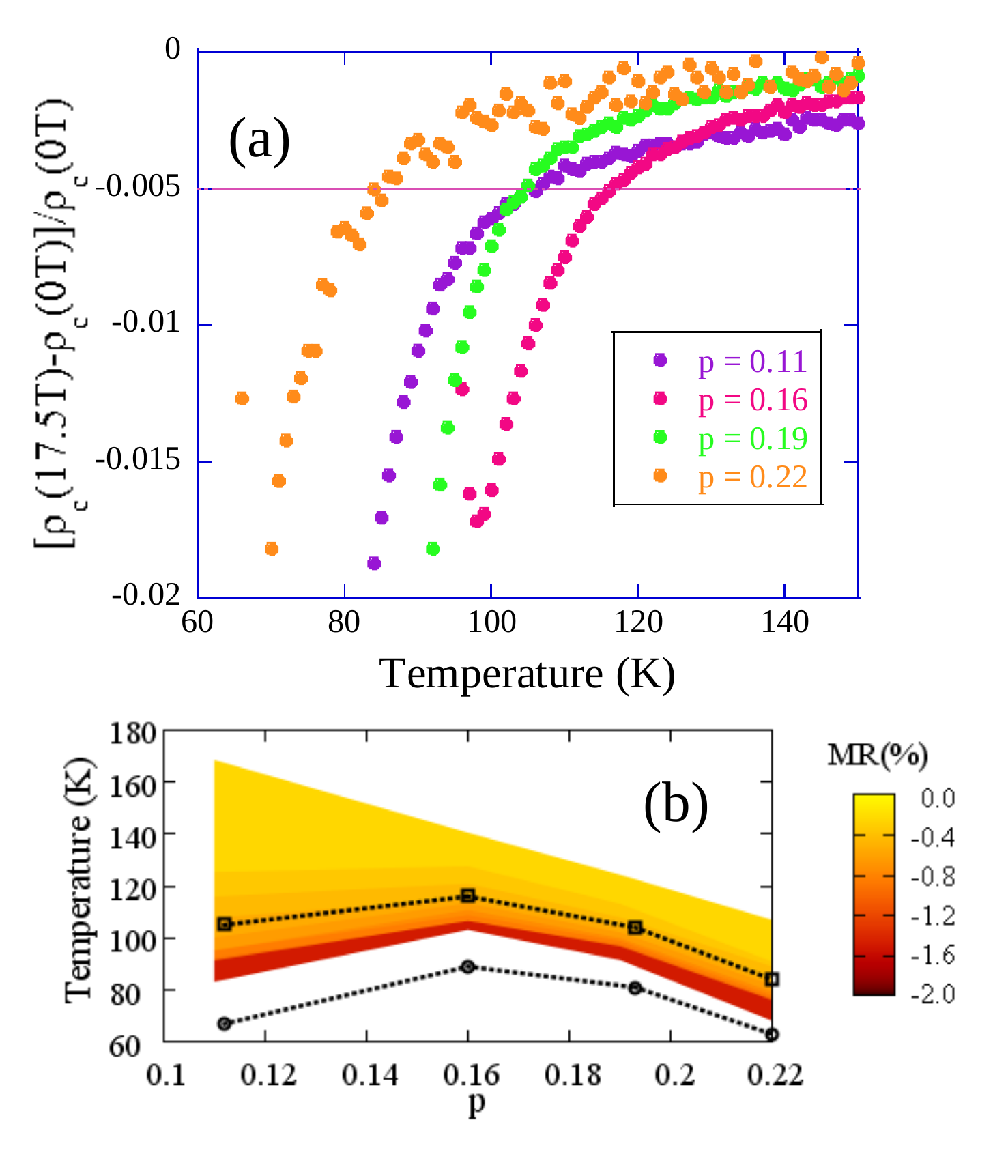}
\caption{\label{f3}(Color online) (a) Temperature dependence of the out-of-plane MR at 17.5 T, ($\rho_{c}$(17.5 T)-$\rho_{c}$(0 T))/$\rho_{c}$(0 T), for variously doped samples. The horizontal line indicates the MR value adopted to determine $T_{scf}{\rho_{c}}$. (b) A contour plot for the temperatures, corresponding to different out-of-plane MR values at 17.5 T vs. hole doping level $p$. Solid squares and solid circles represent $T_{scf}{\rho_{c}}$ and $T_{c}$, respectively.}
\end{figure}

\begin{figure}
\includegraphics[width=85mm]{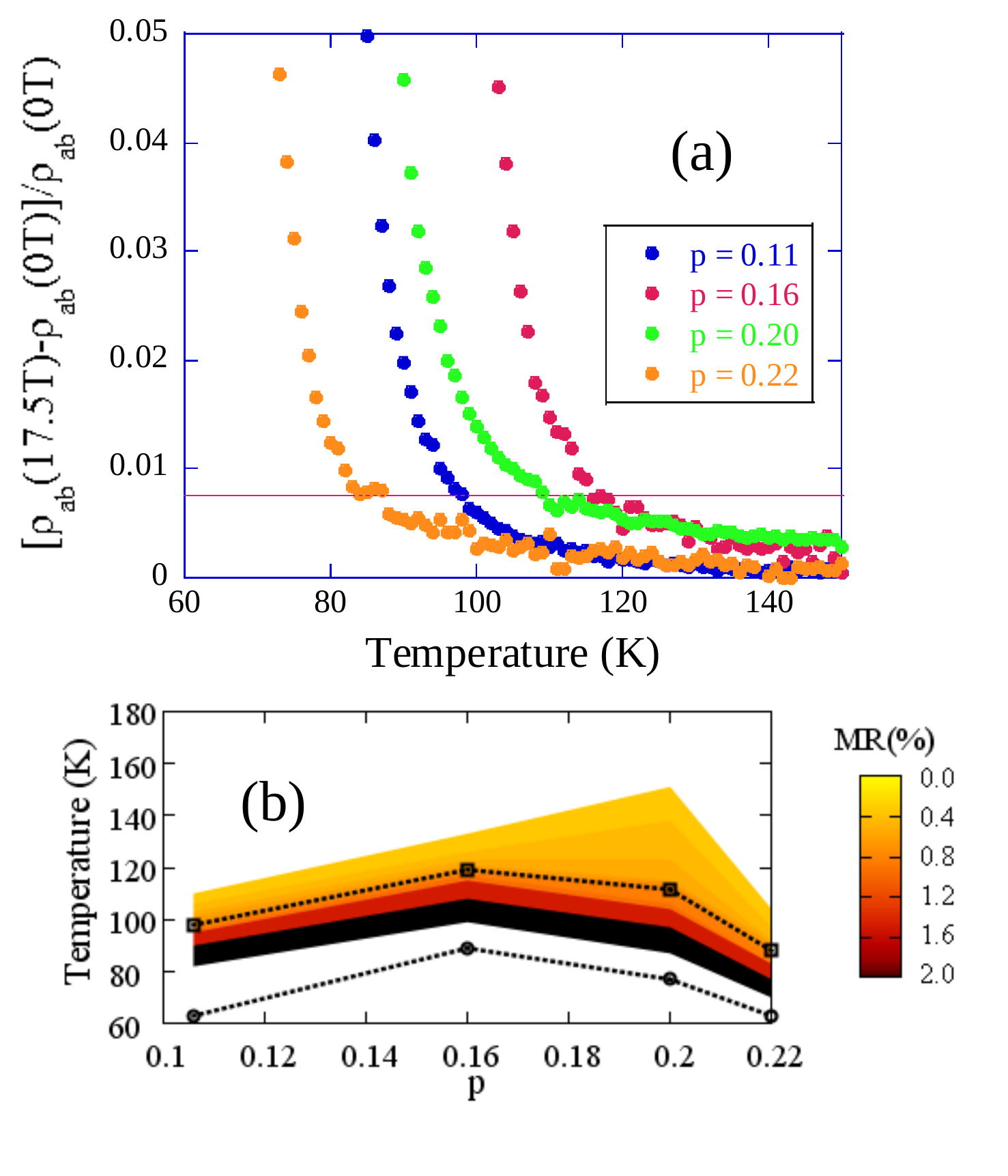}
\caption{\label{f4}(Color online) (a) Temperature dependence of the in-plane MR at 17.5 T, ($\rho_{ab}$(17.5 T)-$\rho_{ab}$(0 T))/$\rho_{ab}$(0 T), for variously doped samples. A horizontal line indicates the MR value adopted to determine $T_{scf}{\rho_{ab}}$. (b) A contour plot for the temperatures, corresponding to different in-plane MR values at 17.5 T vs. hole doping level $p$. Solid squares and solid circles represent $T_{scf}{\rho_{ab}}$ and $T_{c}$, respectively.}
\end{figure}

Figure \ref{f3} (a) shows the temperature dependence of the out-of-plane MR at 17.5 T, ($\rho_{c}$(17.5 T)-$\rho_{c}$(0 T))/$\rho_{c}$(0 T), for the samples shown in Fig. \ref{f1}. We have observed a small negative MR below $T^*_{\rho_{c}}$, which agrees with the previous reports~\cite{t.shiba,lav}. Therefore, the out-of-plane MR is negative in the temperature region shown, except in close vicinity to $T_c$ where the AL superconductive fluctuation contribution dominates. The negative MR increases rapidly with approaching $T_{c}$ for all doping levels. In order to view the situation more clearly, a contour plot for the temperatures, corresponding to different out-of-plane MR values, is shown as a function of doping level $p$ (Fig. \ref{f3}(b)). It is noticeable that the temperature region at which the large negative MR is observed is concentrated near $T_{c}$. Figure \ref{f4} (a) shows the temperature dependence of the in-plane MR at 17.5 T, ($\rho_{ab}$(17.5 T)-$\rho_{ab}$(0 T))/$\rho_{ab}$(0 T), for the samples shown in Fig. \ref{f2}. For the in-plane cases, MR is always positive. For all doping levels, a small MR is observed at temperatures higher than 130 K. This is assigned as the normal-state MR that is proportional to the square of the Hall angle~\cite{harris,semba}. The amplitude of the normal state MR reaches its maximum at a slightly overdoped state ($p$ = 0.20). When the temperature is decreased, the positive MR rapidly increases with approaching $T_{c}$. Figure \ref{f4} (b) shows a contour plot for the temperatures, corresponding to different in-plane MR values as a function of the doping level $p$. A large MR is observed near $T_{c}$. From Figs. \ref{f3}(b) and \ref{f4}(b), it is found that the temperature regions at which the large MR's are observed coincide. Since superconductive fluctuation is expected to lead to positive in-plane MR but to negative out-of-plane MR in extremely 2D high-$T_{c}$ cuprates~\cite{varla}, we have determined that these temperature zones are superconductive fluctuation regions. To estimate the onset temperatures for the rapid growth of negative MR, $T_{scf}{\rho_{c}}$, and positive MR, $T_{scf}{\rho_{ab}}$, critical MR values of -0.005 and 0.0075 have been chosen from Fig. \ref{f3}(a) and Fig. \ref{f4}(a), respectively. $T_{scf}{\rho_{c}}$ is estimated as 105, 116, 105, and 84 K for $p$ = 0.11, 0.16, 0.19, and 0.22, respectively. These $T_{scf}{\rho_{c}}$ values are plotted in Fig. \ref{f3}(b). $T_{scf}{\rho_{ab}}$ are estimated as 98, 117, 109, and 84 K for $p$ = 0.11, 0.16, 0.20, and 0.22, respectively. These $T_{scf}{\rho_{ab}}$ values are plotted in Fig. \ref{f4}(b).

\begin{figure}
\includegraphics[width=85mm]{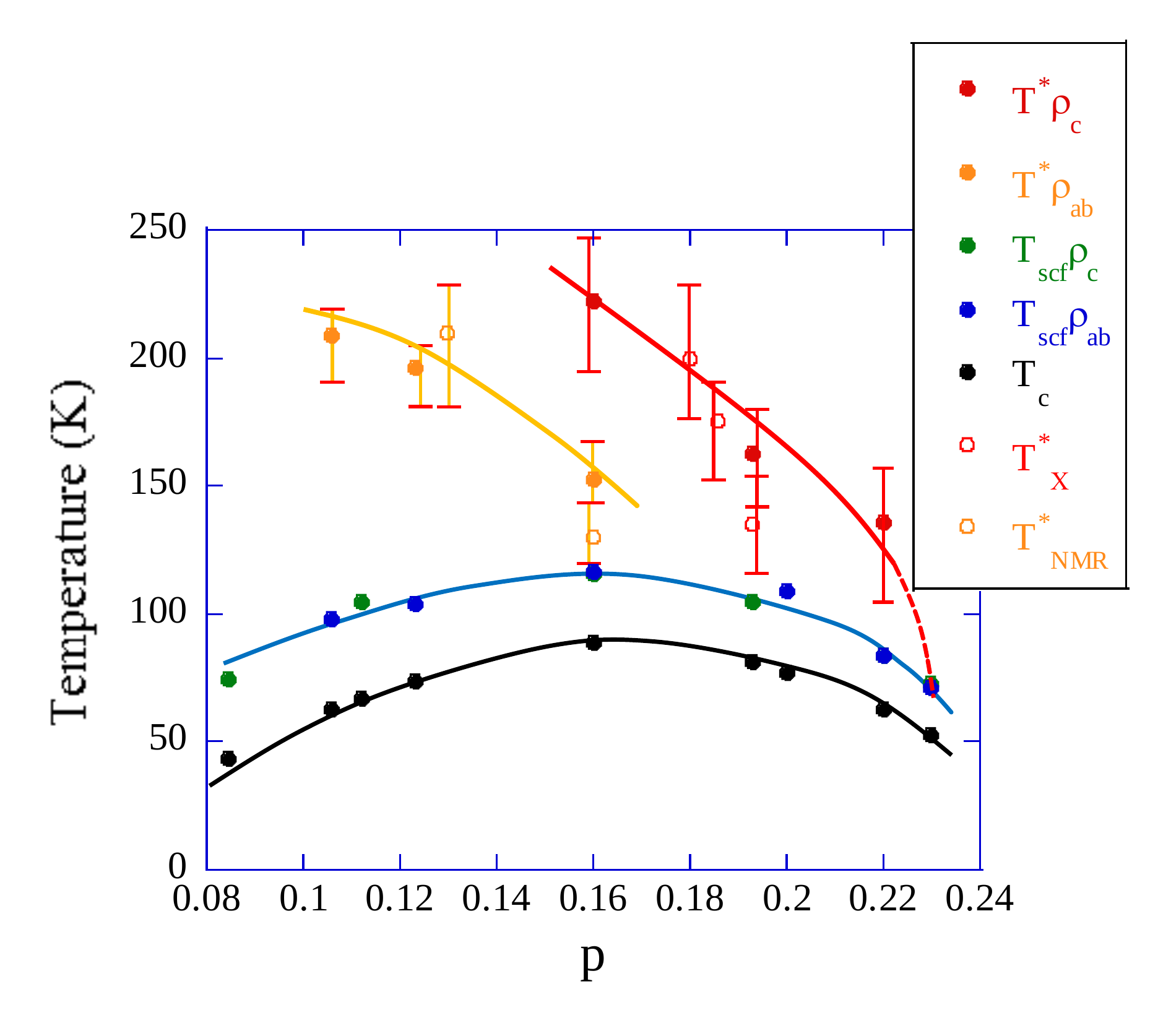}
\caption{\label{f5} (Color online) Characteristic temperatures vs. hole doping level $p$. Closed circles represent characteristic temperatures obtained in this study (some data points are not shown in the text). Open circles of $T^*_{\chi}$ and $T^*_{NMR}$ are replotted from Ref. 13 and Ref. 34, respectively. The $p$ values have been estimated from the empirical $T_c$ vs. $p$ relation.}
\end{figure}

The $T^*_{\rho_{c}}$, $T^*_{\rho_{ab}}$, $T_{scf}{\rho_{c}}$, $T_{scf}{\rho_{ab}}$, and $T_c$ obtained together with several other characteristic temperatures are plotted as a function of a hole doping level $p$ in Fig. \ref{f5}. $T_{scf}{\rho_{c}}$ and $T_{scf}{\rho_{ab}}$ almost coincide for each $p$. These results assure that the large MR in both directions originates from superconductive fluctuation.  The dominant DOS fluctuation effect and the dominant AL fluctuation effect are responsible for the negative out-of-plane MR and positive in-plane MR, respectively. $T_{scf}$ almost completely tracks the $T_c$ 'dome', although a tendency for narrowing of the fluctuating regime is seen closer to overdoping. On the other hand, $T^*$ ($T^*_{\rho_{c}}$) is higher than $T_{scf}$ in the underdoped region. It monotonically decreases upon doping and comes very close to $T_{scf}$ in the overdoped state ($p$ = 0.22). Below the high temperatures
of $T^*_{\rho_{c}}$, $\rho_{c}$ first shows a typical upturn and a small negative MR due to the opening of the pseudogap. The effect of the opening of the superconducting gap is added to $\rho_{c}$ at temperatures near $T_{c}$.

$T^*_{\rho_{c}}$ and $T^*_{\rho_{ab}}$ are different for each $p$. Nevertheless, they are thought to have the same origin, {\it i.e.}, the pseudogap opening, and thus to be the same temperature in principle. The difference in appearance may arise for the following reasons. From the "cold spot model"~\cite{iof} proposed by Ioffe and Millis, the carriers around the ($\pi/a$, 0) direction are exposed to strong scattering (hot spots), while those around the($\pi/2a$, $\pi/2a$) direction are not scattered very much (cold spots). Then, carriers around the cold spots mainly contribute to $\rho_{ab}$ while those around the hot spots, since the hopping probability in the $c$-axis direction is at the maximum here, mainly contribute to $\rho_{c}$. When the pseudogap opens up at $T^*_{\rho_{c}}$ around the hot spots, $\rho_{c}$ gradually increases with decreasing temperatures directly reflecting the effect of DOS depletion by the pseudogap, while $\rho_{ab}$ may practically not be affected. With further decreasing temperatures below $T^*_{\rho_{ab}}$, $\rho_{ab}$ is affected by the pseudogap opening through a suppression of carrier scattering, which is caused by diminution of phase space available for the scattered electrons. In fact, as stated in the introduction, $T^*_{\rho_{c}}$ coincides with $T^*_{\chi}$~\cite{wata2}, below which the magnetic susceptibility gradually decreases. Since the magnetic susceptibility also reflects DOS, this coincidence assures that the pseudogap opens below these temperatures. On the other hand, $T^*_{\rho_{ab}}$ coincides with a characteristic temperature of $T^*_{NMR}$~\cite{ishida} (denoted as $T^*$ in Ref. 34), below which $(T_{1}T)^{-1}$ decreases from the high temperature Curie-Weiss behavior. Since $T^*_{\rho_{ab}}$ and $T^*_{NMR}$ mainly probe carrier scattering, this coincidence indicates that carrier scattering decreases below these temperatures. All these results assure that $T^*_{\rho_{c}}$ directly represents the pseudogap opening through DOS depletion, while $T^*_{\rho_{ab}}$ represents the pseudogap opening indirectly through suppression of carrier scattering. That is, the difference between $T^*_{\rho_{c}}$ and $T^*_{\rho_{ab}}$ may originate in their different sensitivities to the pseudogap opening. It should be noted that $T_{scf}$ is apparently different not only from $T^*_{\rho_{c}}$ but also from $T^*_{\rho_{ab}}$. 

\begin{figure*}
\includegraphics[width=160mm]{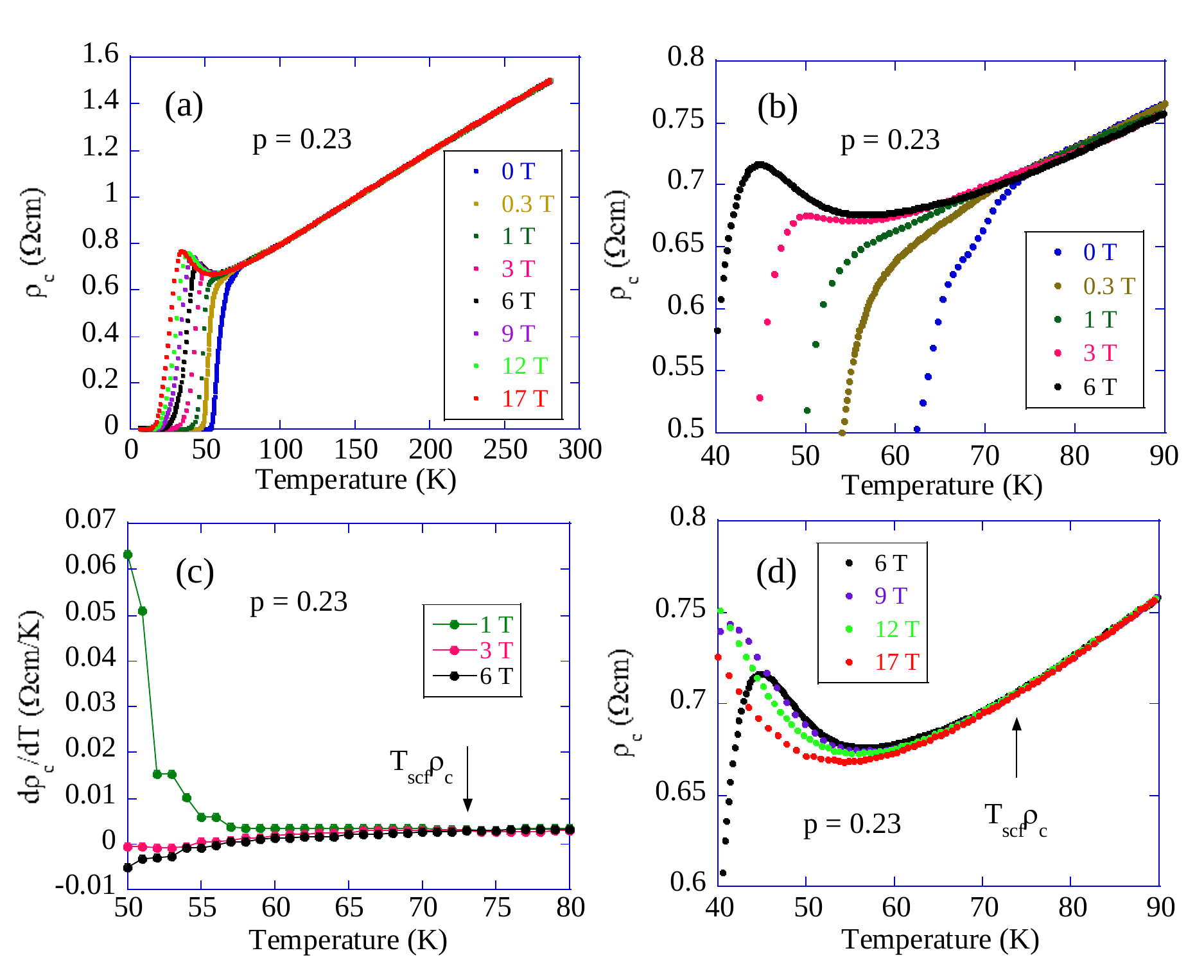}
\caption{\label{f6} (Color online) (a) Out-of-plane resistivity $\rho_c(T)$ for the heavily overdoped ($p$ = 0.23) Bi$_{2}$Sr$_{2}$CaCu$_{2}$O$_{8+\delta}$ single crystal in various magnetic fields. (b) An expanded plot of (a) around $T_c$ for low fields (0 - 6 T). (c) Temperature derivative of $\rho_c$ vs. temperature under applied magnetic fields of 1, 3, and 6 T.  Arrow indicates the temperature $T_{scf}{\rho_{c}}$, below which $d\rho_{c}/dT$ decreases with increasing magnetic field. (d) An expanded plot of (a) around $T_c$ for high fields(6 - 17 T). The arrow indicates the temperature $T_{scf}{\rho_{c}}$ below which $\rho_c$ shows a marked negative magnetoresistance (MR).}
\end{figure*}

To explore the relationship between the pseudogap and superconductivity in detail, a more overdoped state is investigated. Figure \ref{f6} (a) shows the temperature dependence of the out-of-plane resistivity $\rho_{c}$ for a heavily overdoped ($p$ = 0.23) sample under various magnetic fields. At zero field, $\rho_{c}$ exhibits metallic behavior down to $T_c$ (= 53 K). Figure \ref{f6} (b) shows an expanded plot around $T_c$ for low magnetic fields (0--6 T). The zero-field data are affected by a trace of some minor superconductive phases ($T_c$ $\approx$ 70 and 105 K), which are hard to avoid. The effect seems, however, to be suppressed at magnetic fields of 1 T and above. Note that $\rho_{c}$ is still metallic at 1 T. When the applied magnetic field exceeds 1 T, $\rho_{c}$ shows a slight upturn upon cooling. To see its onset clearly, $\rho_{c}$ was differentiated with respect to temperature. Figure \ref{f6} (c) shows the results. When the applied field is 1 T, $d\rho_{c}/dT$ is almost constant above 60 K, whereas when the field is increased to 3 and 6 T, it shows a slight downward deviation below $T_{scf}{\rho_{c}}$ (= 73 K). Thus, this temperature is identified as the onset temperature for the upturn of $\rho_{c}$. When the magnetic field exceeds 6 T, the typical negative MR appears in $\rho_{c}$ below the same temperature, $T_{scf}{\rho_{c}}$ = 73 K [Fig. \ref{f6} (d)]. We have observed similar behavior previously~\cite{wataC} and attributed it to the pseudogap opening. We have considered~\cite{wataC} that, because the pseudogap opens just below $T_c$, the upturn has been hidden by the onset of superconductivity at zero field, but it has emerged along with the suppression of superconductivity with the application of high magnetic fields. 

\begin{figure}
\includegraphics[width=85mm]{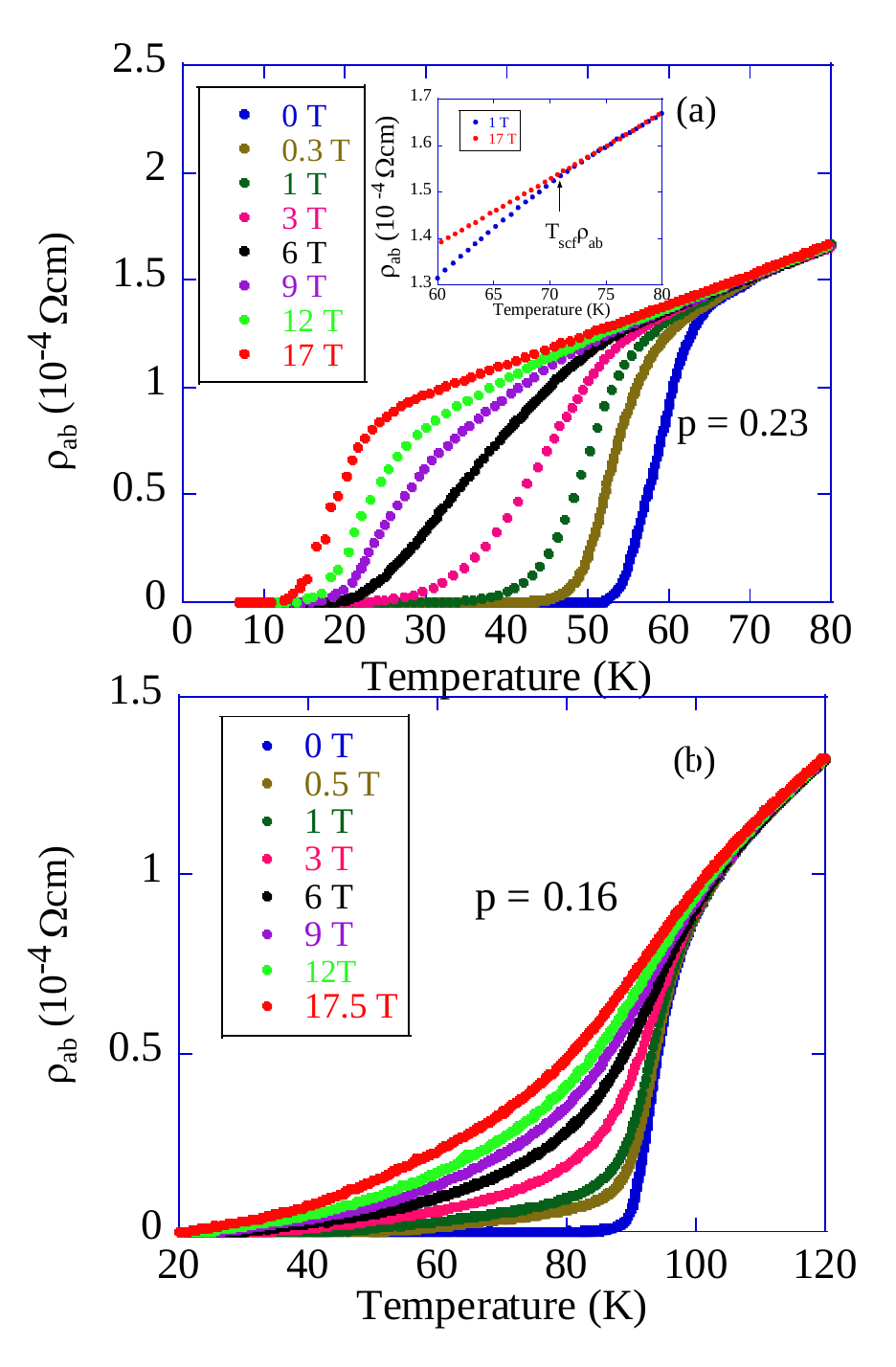}
\caption{\label{f7} (Color online) (a) In-plane resistivity $\rho_{ab}(T)$ for the heavily overdoped ($p$ = 0.23) Bi$_{2}$Sr$_{2}$CaCu$_{2}$O$_{8+\delta}$ single crystal in various magnetic fields. The data with  1 and 17 T is shown in the inset with the scale expanded for a better view around $T_c$. The arrow in the inset indicates the temperature $T_{scf}{\rho_{ab}}$ below which $\rho_{ab}$ shows a marked positive magnetoresistance (MR). (b) In-plane resistivity $\rho_{ab}(T)$ for the optimally doped ($p$ = 0.16) Bi$_{2}$Sr$_{2}$CaCu$_{2}$O$_{8+\delta}$ single crystal (same sample as shown in Fig. \ref{f2} (b)) in various magnetic fields.}
\end{figure}

However, the upturn and the associated negative MR in $\rho_{c}$ can be expected also because of the superconductive DOS fluctuation effect if the system is extremely 2D~\cite{varla}. To clarify this point, the in-plane resistivity $\rho_{ab}$ for the same heavily overdoped ($p$ = 0.23) sample was measured in several magnetic fields. Figure \ref{f7} (a) shows the results. As stated above for $\rho_{c}$, the zero-field data are affected by the existence of minor superconductive phases. Therefore, $T_{scf}{\rho_{ab}}$ was estimated using the data for 1 and 17 T. The inset of Fig. \ref{f7} (a) shows an enlarged view around $T_c$. Marked positive MR is observed below $T_{scf}{\rho_{ab}}$ (= 71 K). $T_{scf}{\rho_{c}}$ and $T_{scf}{\rho_{ab}}$ almost coincide. This result suggests that the pseudogap-like behavior of $\rho_{c}$ [Fig. \ref{f6} (b) and (d)] is caused by the superconductive fluctuation effect. 

It is noticeable in Fig. \ref{f7} (a) that the superconducting transition curve is fan-shaped at low fields, whereas it becomes parallel-shift-like for fields above 6 T. This feature is similar to that observed for stripe-ordered La-214 systems with a hole concentration near 1/8 ($p$ = 0.125) per Cu~\cite{adachi}. This feature was also observed recently for an overdoped La-214 sample with a hole concentration near 1/4 ($p$ = 0.25)~\cite{miya}. These features have been attributed to a magnetic-field-induced stabilization of the stripe order~\cite{adachi} suggested by neutron scattering measurements of the La-214 system~\cite{lake}. A magnetic-field-induced charge-stripe order for YBa$_{2}$Cu$_{3}$O$_{y}$ was also  reported recently~\cite{wu}. To consider the reason for this anomalous $\rho_{ab}$ behavior [Fig. \ref{f7} (a)], the superconducting transition curves for the optimally doped sample ($p$ = 0.16) under various magnetic fields are shown for comparison in Fig. \ref{f7} (b). In contrast to the data in Fig. \ref{f7} (a), those in Fig. \ref{f7} (b) show typical broadening under all the fields. Note that the fan-shaped feature of the transition curve at low fields shown in Fig. \ref{f7} (a) is weak compared with that in Fig. \ref{f7} (b). This reflects the fact that the 2D nature of the heavily overdoped sample is weak. Indeed, the anisotropy parameter $\gamma$ determined by the zero field resistivity ratio ($\gamma = \sqrt{\rho_{c}/\rho_{ab}}$) at $T_{scf}$ is 70 in this heavily overdoped state, which is very small compared with the value of 230 for the optimally doped state. On the other hand, an extreme fan-shaped feature is observed for fields above 6 T for temperatures around 30--50 K, indicating that the system is extremely 2D. Note also that the resistivity at high fields in this temperature region approaches the normal state values. This indicates that the excess conductivity due to superconductive fluctuation is very small, implying that the superfluid density $\rho_{s}$ is reduced under high magnetic fields. These results are consistent with the assumption that magnetic-field-induced stabilization of the stripe order has occurred. Because the orientation of the stripe order rotates by $\pi$/2 from one layer to the next, the formation of the stripe order is thought to electronically decouple the CuO$_{2}$ planes, leading to 2D superconductivity~\cite{li}. In fact, a 2D superconducting transition (a Berezinskii--Kosterlitz--Thouless transition) has been reported for a stripe-ordered La$_{1.875}$Ba$_{0.125}$CuO$_{4}$~\cite{li}. $\rho_{s}$ may also be reduced by the formation of the stripe order. At low temperatures (around 20 K), a sharp resistive transition is observed. This is probably due to the strong pinning of the vortex. Consequently, the parallel-shift-like behavior observed at high fields may be the combined effect of the broadening at high temperatures due to the extreme 2D nature and the sharp drop at low temperatures due to vortex pinning. From these facts, we conclude that magnetic-field-induced stabilization of the stripe order occurred above 6 T, causing a dimensional crossover from weak to extreme 2D in this heavily overdoped state. To the best of our knowledge, this result is the first observation of the development of stripe order in overdoped Bi-2212. This reveals that the tendency toward stripe order is general in high-$T_c$ cuprates, and it appears at a hole concentration not only of 1/8 but also of 1/4 per Cu. 
 
Then, it is more natural to consider that the pseudogap-like behavior of $\rho_{c}$ for the heavily overdoped ($p$ = 0.23) sample is caused by the superconductive fluctuation effect. At low fields of 1 T, because the 2D nature of the system is weak, $\rho_{c}$ exhibits metallic behavior. When the magnetic field is increased to 3 T, the 2D nature is increased, causing a slight upturn in $\rho_{c}$. At high fields of more than 6 T, because the system is extremely 2D owing to the stabilized stripe order, $\rho_{c}$ shows a steep upturn and the associated large negative MR [Fig. \ref{f6} (d)]. This high field behavior may be attributed to the fact that the superconductive DOS fluctuation contribution is dominant over the AL contribution for $\rho_{c}$ in the extreme 2D system~\cite{varla}. Here, we would like to correct our previous assertions~\cite{wataC}. 

Finally, let us discuss the implication of the obtained results. From an underdoped ($p$ = 0.11) to an overdoped ($p$ = 0.22) state, it was found that $T^*$ differs from $T_{scf}$. This observation agrees with many other studies~\cite{lav,kudo,tajima1,chen,ong}. Furthermore, $T^*$ was recently shown~\cite{tajima} to exist even for the heavily Zn-doped non-superconducting YBa$_{2}$(Cu$_{1-x}$Zn$_{x}$)$_{3}$O$_{y}$ by using the $c$ axis optical spectra. As pointed out in Ref. 43, these results indicate that the pseudogap is not a simple precursor of superconductivity. On the other hand, in our heavily overdoped ($p$ = 0.23) sample, $T^*$ was not found above $T_{scf}$. If the pseudogap existed above $T_{scf}$, it would be detected as the upturn in $\rho_{c}$ even under a magnetic field. One should remember that the pseudogap is robust against the magnetic field compared with the superconductivity~\cite{kra}. Then, two scenarios arise for $T^*$ of this heavily overdoped state. One is that $T^*$ does exist, but it just equals $T_{scf}$ or exists below it. In this case, the pseudogap is distinct from superconductivity and may coexist with it. Unfortunately, however, it is difficult to separate the signatures of the pseudogap from the superconductive fluctuation below $T_{scf}$. Recent in-plane MR measurements of optimally doped YBa$_{2}$Cu$_{3}$O$_{y}$ using pulsed high magnetic fields indicated~\cite{alloul} that $T^*$ has been found below $T_{scf}$. Another possibility is that $T^*$ never exists down to absolute zero temperature. In this case, we suppose that the pseudogap in less-doped states has eventually changed to a superconducting gap in this heavily overdoped state. Out-of-plane MR measurements of Tl$_{2}$Ba$_{2}$CuO$_{6+x}$ (Tl-2201)~\cite{elbaum} have shown that $T^*$ has not been observed in the extremely overdoped state near the edge of the superconducting ``dome.'' The authors of that report argued that quantum critical fluctuations that destabilize the pseudogap are responsible for the high-$T_c$ superconductivity. Therefore, to resolve this issue, it is crucial that we determine whether or not the pseudogap opens after the superconducting gap opens.  

\section{CONCLUSIONS}

The resistivities $\rho_{ab}(T)$ and $\rho_{c}(T)$ of Bi-2212 single crystals were systematically measured under a wide range of doping conditions in various magnetic fields. Large positive in-plane MR and negative out-of-plane MR were observed near $T_c$. Those temperature zones were found to coincide. Because, for extremely 2D systems,
superconductive fluctuation is expected to lead to positive in-plane MR but negative out-of-plane MR~\cite{varla}, we determined that these temperature zones are superconductive fluctuation regions. It has been found that, from an underdoped to an overdoped state, the onset temperature of superconductive fluctuation, $T_{scf}$, is below the pseudogap opening temperature $T^*$. This result indicates that the pseudogap is not a simple precursor of high-$T_c$ superconductivity. On the other hand, a heavily overdoped state ($p$ = 0.23) was also revealed. In this doping state, we could not find $T^*$ above $T_{scf}$. It has been shown that the pseudogap is interpreted differently depending on whether or not $T^*$ exists below $T_{scf}$. Therefore, we propose that clarifying this point is important to a comprehensive understanding of the role of the pseudogap in high-$T_c$ superconductivity.

\begin{acknowledgment}

We thank L. Taillefer, E. Milani, T. Fujii, A. Matsuda, M. Suzuki, H. J. Im, J. Goryo, and S. Adachi (SRL) for helpful input. The magnetoresistance measurements were performed at the High Field Laboratory for Superconducting Materials, Institute for Materials Research, Tohoku University.


\end{acknowledgment}



\end{document}